\documentclass[aps,twocolumn,groupedaddress]{revtex4}
\usepackage{graphicx}
\usepackage{float}

\begin{document}

\title{Enhanced millimeter wave transmission through subwavelength hole
arrays}

\author{M. Beruete and M. Sorolla}
\affiliation{Departamento de Ingenier\'{\i}a El\'ectrica y
Electr\'onica, Universidad P\'ublica de Navarra, E-31006 Pamplona,
Spain.}
\author{I. Campillo and J.S. Dolado}
\affiliation{Labein Centro Tecnol\'ogico, Cuesta de Olabeaga 16,
E-48013, Bilbao, Spain.}
\author{L. Mart\'{\i}n-Moreno}
\affiliation{Departamento de F\'{\i}sica de la Materia
Condensada,ICMA-CSIC, Universidad de Zaragoza, E-50009 Zaragoza,
Spain.}
\author{J. Bravo-Abad and F.J. Garc\'{\i}a-Vidal}
\affiliation{Departamento de F{\'{\i}}sica Te\'orica de la Materia
Condensada, Universidad Aut\'onoma de Madrid, E-28049 Madrid,
Spain.}

\begin{abstract}
In this letter, we explore, both experimentally and theoretically,
the existence in the millimeter wave range of the phenomenon of
extraordinary light transmission though arrays of subwavelength
holes. We have measured the transmission spectra of several
samples made on aluminum wafers by using an AB Millimetre$^{TM}$
Quasioptical Vector Network Analyzer in the wavelength range
between $4.2$mm to $6.5$mm. Clear signals of the existence of
resonant light transmission at wavelengths close to the period of
the array appear in the spectra.
\end{abstract}

\maketitle 



The discovery of the phenomenon of extraordinary optical
transmission (EOT) observed in two-dimensional (2D) arrays of
subwavelength holes perforated in optically thick metallic films
\cite{Ebbesen}, has opened up the possibility of using
subwavelength apertures for a variety of optoelectronic
applications. A previous theoretical work\cite{LMM} on Ebbesen's
experiment assigned the EOT phenomenon to the excitation of
surface electromagnetic (EM) modes occurring on corrugated metal
surfaces. Furthermore, these modes (and EOT) were found to appear
even in a simpler model where the metal was treated as a perfect
conductor \cite{LMM,Pendry04}. These surface leaky modes are
similar to the ones appearing in perfectly conducting sinusoidal
gratings \cite{Maystre}. As the perfect conductor approximation
should be even more valid for larger wavelengths, the previously
cited work \cite{LMM} pointed out to the possibility of the
existence of EOT in other ranges of the EM spectrum. Moreover,
very recently, there have been some experimental studies of EOT in
the THz regime in doped semiconductors \cite{Rivas} and in metals
\cite{theothers} that seem to suggest that EOT is also present in
this frequency regime.

Here we move a step further by studying, both experimentally and
theoretically, the transmission of EM radiation through 2D arrays
of subwavelength holes in the millimetric wave range. In order to
carry out our analysis, several prototypes have been fabricated in
aluminum wafers of different thicknesses ($w$), ranging from
$0.5$mm to $4$mm. All square arrays ($31 \times 31$) have a
lattice constant ($d$) of $5$mm and two different hole radius
($R$) are considered: $1.25$ and $1$mm (see Fig.1a).

\begin{figure}[h]
\begin{center}
\includegraphics[width=8 cm]{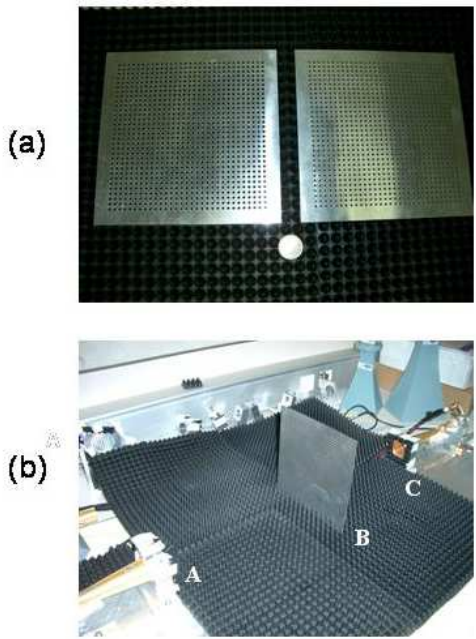}
\end{center}
\vspace*{-0.8cm} \caption{(a) Photograph of two of the six samples
analyzed in the experiment (left image: $R=1.25$mm, $w=0.5$mm and
right image: $R=1$mm, $w=0.5$mm). (b) Photograph of the
experimental setup with: (A) corrugated horn antenna acting as
source of millimeter waves (B) sample and (C) receptor antenna.}
\end{figure}

It is important to note here that, before Ebbesen experiment,
there were experimental studies of transmission of light through
arrays of holes in the far infrared \cite{Ulrich}, mid infrared
\cite{Rhoads} and infrared \cite{Keilmann} ranges. However, these
previous experiments were performed for hole sizes and lattice
constants ($d$) such that $d$ was smaller than the cut-off
wavelength ($\lambda_c$). EOT appears essentially $\lambda=d$, but
when the modes inside the hole are evanescent, i.e., when $d >
\lambda_c$.

Let us first discuss the theoretical predictions for the
transmittance spectra given by the framework described
in\cite{LMM} for the study of EOT in the optical range. Within
this formalism, we consider a metal film perforated by an {\bf
infinite} 2D square array of holes. As aluminum behaves as a
quasi-perfect conductor in the millimeter wave regime, we have
simplified our formalism by considering perfect metal boundary
conditions (PMBC) at all interfaces forming the structure. Within
the PMBC approximation, this theoretical framework is rigorous,
being equivalent to the one developed some time ago for studying
inductive grids \cite{McPhedran}.

\begin{figure}[h]
\begin{center}
\includegraphics[width=7.3 cm]{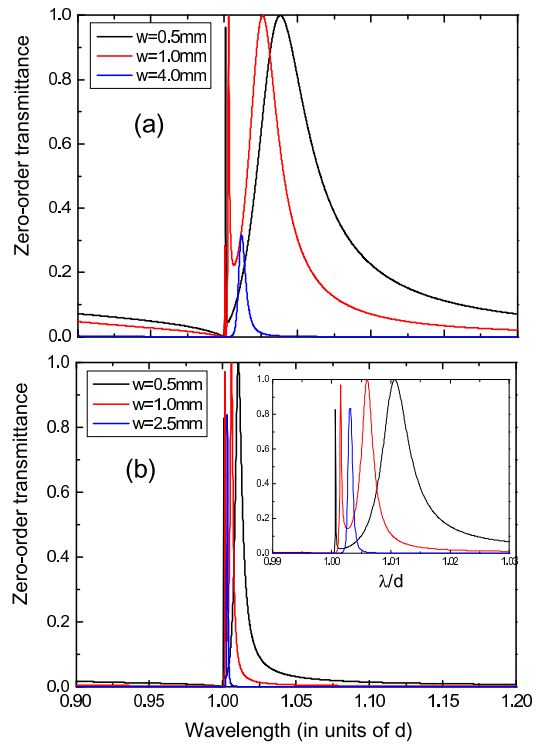}
\end{center}
\vspace*{-0.8cm} \caption{Theoretical zero-order transmittance
spectra corresponding to different infinite hole arrays with
$R=1.25$mm and three different thicknesses $w$'s in panel
(a)($w=0.5$mm (black), $w=1$mm (red) and $w=4.0$mm (blue)) and
$R=1$mm with three $w$'s in panel (b) ($w=0.5$mm (black), $w=1$mm
(red) and $w=2.5$mm (blue)). Inset to panel (b) represents the
same physical quantity but in a smaller range of wavelengths.}
\end{figure}

In Fig. 2 we show our numerical simulations for the zero-order
transmittance spectra of infinite arrays of holes corresponding to
the six samples fabricated. In all calculations we show in this
paper we assume that a normal incident plane wave is impinging at
the perforated metal film. Panel (a) displays the cases with
$R=1.25$mm ($\lambda_c= 0.85 d$) and three different thicknesses
and in panel (b) the corresponding three transmittance spectra
with $R=1$mm ($\lambda_c= 0.68 d$) are shown. In the region
$\lambda/d \approx 1$, calculations rendered in panels 2a and 2b
predict the appearance of EOT resonances. For each of the thinner
samples considered ($w=0.5$mm and $w=1.0$mm), the two surface EM
resonances excited at the two surfaces of the metallic film are
coupled, leading to two transmission peaks that reach $100 \%$
\cite{LMM}. However, for the thicker samples analyzed ($w=4.0$mm
for $R=1.25$mm and $w=2.5$mm for $R=1$mm), this EM coupling is
negligible and only one transmission peak appears in the spectra.

Transmission through our samples is measured by using an AB
Millimeter$^{TM}$ Quasioptical Vector Network Analyzer in the
frequency range between $40$ to $110$GHz. In Fig.1b we show a
photograph of the experimental set-up. A vertically polarized pure
gaussian beam is generated by a corrugated horn antenna (A). This
beam propagates up to the sample (B) that is located at $166$cm to
the antenna. The diameter of the beam waist at the sample location
is around $50$cm at the wavelength range of interest. In this way,
the illumination of the hole arrays is rather uniform. The
transmitted beam is finally collected into a horn antenna (C) that
is placed at $105$cm from the sample. The samples are embedded
into a sheet of millimeter wave absorbing material (not shown in
Fig.1b for illustrative purposes) in such a way that any possible
diffracted beam generated by the edges of the samples is absorbed
by the sheet and not collected by the receiver antenna (C).

Fig.3 shows experimental transmission spectra obtained for the six
different samples analyzed. We represent the collected
transmission power, $T$ (normalized to the collected power when no
sample is present) as a function of wavelength, when the holes
have radius $R=1.25$mm (top panel, full lines) and when $R=1$mm
(bottom panel, full lines). In the case of $R=1.25$mm and
$w=0.5$mm (black curve in panel a), the transmission at resonance
(located at $\lambda$ slightly larger than $d$) can be as large as
$95 \%$ although the holes only occupy $20 \%$ of the unit cell.
For $R=1.25$mm and $w=1$mm (red full curve), the transmission
resonance also appears, reaching $65 \%$ at maximum. This kind of
transmission resonances is also present in the thinner samples
($w=0.5$mm and $w=1$mm, see panel b) of the arrays of holes with
smaller radius ($R=1$mm) but the transmittance peaks associated
are much lower than the ones obtained for $R=1.25$mm. For the
thicker films analyzed ($w=4.0$mm for $R=1.25$mm and $w=2.5$mm for
$R=1$mm), the collected power is extremely small and no
fingerprints of transmission resonances are observed. As the
measured transmission resonances appear in a frequency range in
which the holes only support EM evanescent waves, we can safely
conclude that EOT also takes place in the millimeter wave range,
as theory predicted.

\begin{figure}[h]
\vspace{1 cm}
\begin{center}
\includegraphics[width=7.3 cm]{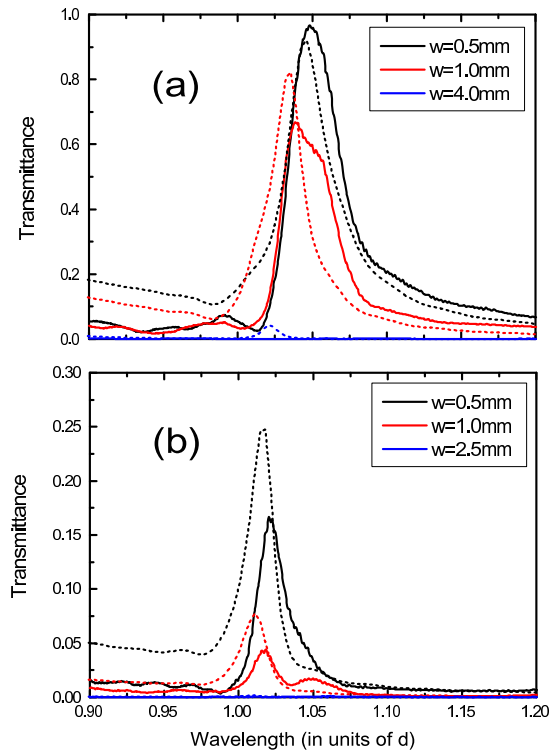}
\end{center}
\vspace*{-0.8cm} \caption{Experimental transmittance spectra (full
lines) and theoretical total transmittance curves (dashed lines)
for the $31 \times 31$ arrays corresponding to the same
geometrical parameters as in Figure 2. Different hole arrays with
$R=1.25$mm in (a) and $R=1$mm in (b) are analyzed.}
\end{figure}

However, there is a strong disagreement between theory and
experiment as regards the absolute value of the transmission
peaks. A possible reason for this disagreement could be originated
by the intrinsic finite size of our arrays ($31 \times 31$). In
order to explore in more detail this possibility, we have applied
a theoretical formalism recently developed in our group that is
able to analyze the optical properties of {\bf finite} collections
of holes drilled in a metallic film \cite{Jorge}. In this method,
first we assume an artificial square supercell of side $L$ in
which the $N \times N$ array of holes is contained ($L
> Nd$). Then, we apply a modal expansion of the EM fields (plane
waves in vacuum regions and TE/TM modes inside the holes
\cite{evanes}) and we match these fields considering PMBC. At the
end of this procedure, a set of linear equations for the expansion
coefficients of the EM-fields at the different holes of our
structure is established. As the considered supercell is
fictitious, we have to take the limit $L \rightarrow \infty$ in
the different terms appearing in this set of linear equations.
Once the expansion coefficients are obtained, the total
transmittance through the 2D subwavelength hole array can be
finally calculated.

In both panels of Fig. 3 (dashed curves), we show the total
transmittance spectra for the six $31 \times 31$ arrays of holes
as obtained with our new theoretical tool. If we compare the
theoretical results for the 31x31 arrays (dashed lines in Fig.3)
with the corresponding ones for infinite arrays (Fig.2) there are
two main changes. Firstly, the very narrow transmission peaks
appearing at $\lambda \approx d$ for the infinite arrays are not
present in the spectra of finite arrays. Secondly, there is a
strong reduction of the transmission peaks when going from
infinite arrays to finite ones; this reduction is more dramatic
for the structures with $R=1$mm than for the arrays with
$R=1.25$mm. These two changes lead to a much better agreement
between theoretical predictions and experimental results (see
Fig.3). This good agreement allows us to state that the strength
of transmission resonances associated to the EOT phenomenon
observed in subwavelength hole arrays is basically controlled by
the size of the arrays.

In conclusion, we have demonstrated that the phenomenon of
extraordinary EM transmission through arrays of subwavelength
holes is also present in the millimeter wave range. Moreover, we
have also shown that one of the key parameters to observe this
phenomenon is the number of periods of the array.

Financial support by the Spanish MCyT under grant BES-2003-0374
and contracts MAT2002-01534 and MAT2002-00139 and by EC project
FP6-NMP4-CT-2003-505699 (``Surface Plasmon Photonics'') is
gratefully acknowledged.


\begin{thebibliography}{}
\bibitem{evanes} As we are working in the subwavelength regime,
considering just the two least decaying modes in each hole is
enough to obtain accurate numerical results.
\end{thebibliography}
\end{document}